\documentclass[prd,twocolumn,superscriptaddress,preprintnumbers,balancelastpage,nofootinbib,floatfix]{revtex4-2}
\usepackage{graphicx}  
\usepackage{dcolumn}   
\usepackage{bm}        
\usepackage{amssymb, amsmath}
\usepackage{slashed}   
\usepackage{lipsum}
\usepackage[usenames,dvipsnames,svgnames,table]{xcolor}
\usepackage{comment}
\usepackage[utf8]{inputenc}
\usepackage{multirow}
\newcommand{\ov}{\overline}

\definecolor{BlueViolet}{rgb}{0.2, 0.00, 0.7}
\definecolor{Blue}{rgb}{0.15, 0.00, 0.9}
\definecolor{lightblue}{rgb}{0.15, 0.35, 0.95}
\definecolor{kitgreen}{rgb}{0,
0.58823 
, 0.50980 
}
\usepackage[
colorlinks=true, linkcolor=lightblue,citecolor=lightblue,urlcolor=kitgreen]{hyperref} 

\newcommand{\eg}{{\em e.g.}}

\usepackage{adjustbox}
\usepackage{tabularx}
\newcolumntype{Y}{>{\centering\arraybackslash}X} 
\usepackage{booktabs} 
\usepackage{colortbl} 
\def\beq#1\eeq{\begin{align}#1\end{align}}

\RequirePackage{xspace}
\def\Bbar    {\kern 0.18em\overline{\kern -0.18em B}{}\xspace}
\def\Bb      {\ensuremath{\Bbar}\xspace}

\newcommand{\bbms}{\bbs\ mixing}

\newcommand{\bbs}{\ensuremath{B_s\!-\!\Bbar{}_s\,}}

\newcommand{\rdst}{\ensuremath{R_{D^{(\ast)}}}\;}
\newcommand{\RDs}{$R_{D^{(*)}}$}

\begin{document}
\widetext
\preprint{P3H--23--046, TTP23--026, CHIBA--EP--263}

\title{Electric Dipole Moments as Probes of \texorpdfstring{\boldmath{$B$}}{B} Anomaly}

\author{Syuhei Iguro}
\email{igurosyuhei@gmail.com}
\affiliation{Institute for Advanced Research, Nagoya University, Nagoya 464--8601, Japan}
\affiliation{Kobayashi-Maskawa Institute for the Origin of Particles and the Universe, Nagoya University, Nagoya 464--8602, Japan}
\affiliation{Institute for Theoretical Particle Physics (TTP), Karlsruhe Institute of Technology (KIT),
Wolfgang-Gaede-Str.\,1, 76131 Karlsruhe, Germany}
\affiliation{ Institute for Astroparticle Physics (IAP),
Karlsruhe Institute of Technology (KIT), 
Hermann-von-Helmholtz-Platz 1, 76344 Eggenstein-Leopoldshafen, Germany}

\author{Teppei Kitahara}
\email{kitahara@chiba-u.jp}
\affiliation{Department of Physics, Graduate School of Science,
Chiba University, Chiba 263--8522, Japan}
\affiliation{CAS Key Laboratory of Theoretical Physics, Institute of Theoretical Physics, Chinese Academy of Sciences, Beijing 100190, China}
\affiliation{Kobayashi-Maskawa Institute for the Origin of Particles and the Universe, Nagoya University, Nagoya 464--8602, Japan}

\begin{abstract}
The measurements of the lepton flavor universality (LFU) in $\mathcal{B}({\,\overline{\!B}} \to D^{(\ast)} l \overline{\nu})$ indicate a significant deviation 
from the standard model prediction at a 3--4 $\sigma$ level,
revealing a violation of the LFU ($R_{D^{(\ast)}}$ anomaly). 
It is known that 
the $R_{D^{(\ast)}}$ anomaly can be easily accommodated by an $SU(2)_L$-singlet vector leptoquark (LQ)  coupled primarily to third-generation fermions, whose existence is further motivated by a partial gauge unification.
In general, such a LQ naturally leads to additional $CP$-violating phases in the LQ interactions.
In this paper, 
we point out that the current $R_{D^{(\ast)}}$ anomaly prefers the $CP$-violating interaction although  $\mathcal{B}({\,\overline{\!B}} \to D^{(\ast)} l\overline{\nu})$ are $CP$-conserving observables. 
The $CP$-violating LQ predicts a substantial size of the bottom-quark electric dipole moment (EDM), the chromo-EDM, and also the tau-lepton EDM.
Eventually at low energy, the nucleon and electron EDMs are radiatively induced.
Therefore, we conclude that the $R_{D^{(\ast)}}$ anomaly with the $SU(2)_L$-singlet vector LQ provides unique predictions:
neutron and proton EDMs with opposite signs and a magnitude of
$\mathcal{O}(10^{-27})\,e\,$cm, and suppressed electron EDM.
Furthermore, we show that a similar EDM pattern is predicted in an $SU(2)_L$-doublet scalar LQ scenario that can accommodate the $R_{D^{(\ast)}}$ anomaly as well.
These EDM signals could serve as crucial indicators in future experiments.
\end{abstract}
\maketitle
\section{Introduction}
\label{sec:introduction}

In the near future, the sensitivities of precision measurements for the elementary particles, particularly the $B$ physics and the electric dipole moments (EDMs), are expected to be improved by an order of magnitude. 
Many kinds of new physics models will undoubtedly be probed through these improvements.

Currently, a significant deviation from the standard model (SM) prediction
has been reported 
by the BaBar, LHCb, Belle, and Belle~II experiments \cite{Lees:2012xj,Lees:2013uzd,Huschle:2015rga,Hirose:2016wfn,Hirose:2017dxl,Abdesselam:2019dgh,Belle:2019rba,Aaij:2015yra,Aaij:2017uff,Aaij:2017deq,LHCb:2023zxo,LHCb:2023cjr,Belle2,Belle-II:2024ami,LHCb2024},
in measurements of the lepton flavor universality (LFU) 
in $\Bb \to D^{(\ast)} l \overline{\nu}$.
Violation of the LFU is represented by 
 \begin{align}
  R_{D^{(*)}} \equiv \frac{\mathcal{B}(\Bb\rightarrow D^{(*)} \tau \ov\nu_\tau)}{\mathcal{B}(\Bb\rightarrow D^{(*)} \ell\ov\nu_\ell)}\,,
\end{align}
where $\ell$ represents an average of the leptons.
The up-to-date world average of the data \cite{HeavyFlavorAveragingGroup:2022wzx,HFLAV2024winter} is  
\beq
R_D^{\rm exp}=0.344 \pm 0.026\,,  \quad R_{D^\ast}^{\rm exp}  = 0.285 \pm 0.012\,,
\eeq
while an up-to-date SM prediction \cite{Bordone:2019guc,Iguro:2020cpg,Bernlochner:2022ywh,Iguro:2024hyk} is 
\beq
R_D^{\rm SM}=0.290\pm 0.003\,,  \quad 
R_{D^\ast}^{\rm SM}  = 0.248 \pm 0.001\,,
\eeq
which implies more than $4\sigma$ level tension.
This $R_{D^{(*)}}$\,anomaly naively suggests the existence of $\mathcal{O}(1)$\,TeV new physics in the $b \to c \tau \ov\nu_\tau$ process, and
various kinds of models have been proposed \cite{London:2021lfn,Iguro:2024hyk}.
One of new physics candidates is an $SU(2)_L$-singlet  vector leptoquark (LQ), dubbed as $U_1$ LQ whose gauge charge is $(\bm{3},\bm{1},2/3)$.
The $U_1$ LQ hypothesis has been widely discussed 
in connection with a partial gauge unification \cite{DiLuzio:2017vat,Bordone:2017bld,Bordone:2018nbg}
as well as  
 the related flavor processes and the LHC phenomenology 
have been studied \cite{Calibbi:2017qbu,Capdevila:2017iqn}.
These new physics predictions will be tested in the ongoing Belle II \cite{Belle-II:2018jsg} and LHCb experiments \cite{Cerri:2018ypt}.
One should note that to avoid the strict constraint from $K_L\to \mu e$ measurements~\cite{Hung:1981pd,Valencia:1994cj}, 
 (elaborate) $U(2)$ flavor symmetries have been considered 
for a successful interpretation of the $R_{D^{(*)}} $ anomaly
\cite{Pomarol:1995xc,Barbieri:1995uv,Barbieri:2011ci,Barbieri:2015yvd}.
In that case, the $U_1$ LQ couples primarily to third-generation fermions.

\begin{figure}[t]
~~~
 \includegraphics[width=.47\textwidth]{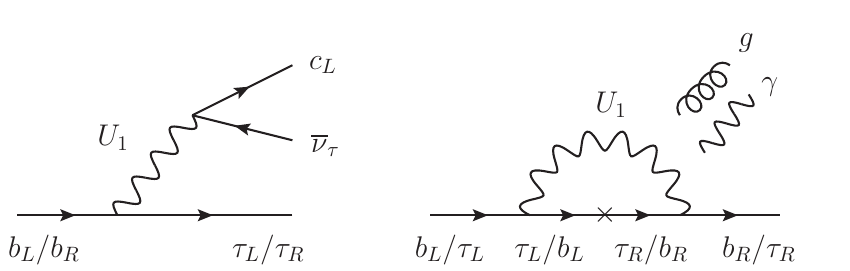}
   \vspace{-.3cm}
\caption{The vector-LQ ($U_1$) contributions to
\rdst (left diagram) and 
the (chromo-) EDMs for 
the bottom-quark and tau-lepton (right diagram).}
    \label{fig:dia}
    \vspace{-.6cm}
\end{figure}
The LQ model naturally brings a $CP$-violating (CPV) phase coming from the rotation matrices to the mass bases of the left- and right-handed quarks and lepton fields that are not aligned in general.
In this paper, it will be clarified that the CPV phase is {\em necessary} to accommodate the \rdst anomaly, 
and this phase also induces the sizable nucleon EDMs at the low energy, which will be 
 testable in the near future (see Fig.~\ref{fig:dia} for the Feynman diagrams).
Although Refs.~\cite{Panico:2018hal,Altmannshofer:2020ywf} investigated the EDMs in the vector-LQ model in light of the \rdst anomaly, they focused on the benchmark point parameters and the necessity of the CPV phase was unclear.

Similar to the vector $U_1$ LQ, 
an $SU(2)_L$-doublet scalar LQ, $R_2$, also produces  a robust correlation between the \rdst anomaly and the nucleon EDMs  \cite{Dekens:2018bci,Gisbert:2019ftm,Babu:2020hun,Becirevic:2022tsj,Kirk:2023fin}.
We will also investigate the $R_2$ LQ scenario 
in a similar way to the $U_1$ LQ scenario.

The paper is organized as follows.
In Sec.~\ref{sec:model}, we introduce a simplified parametrization of the vector $U_1$ LQ scenario.
In Sec.~\ref{sec:EDM}, we obtain the formulae for the relevant observables; EDMs, $R_{D^{(\ast)}}$, and  $B_s \to \tau^+ \tau^-$ in the $U_1$ LQ scenario with several comments. 
In Sec.~\ref{sec:resultU1}, we summarize the result of the $U_1$ LQ scenario. 
Furthermore, in Sec.~\ref{sec:R2}, we investigate the scalar $R_2$ LQ scenario as well.
Finally, we conclude in Sec.~\ref{sec:summary}. 
The additional correlations with the polarization observables are briefly summarized in Appendix~\ref{sec:Appendix_other_bctaunu}.

\section{Vector \texorpdfstring{\bm{$U_1$}}{U1} LQ Scenario}
\label{sec:model}

In the following Secs.~\ref{sec:model}--\ref{sec:resultU1}, we consider a simplified $U_1$ LQ scenario with a $U(2)$ flavor symmetry.
The relevant fermion interactions are  described by 
\begin{align}
\mathcal{L}= \left(\beta_L^{ij} \ov{Q}_i \gamma_\mu P_L L_j + \beta_R^{ij} \ov{d}_i \gamma_\mu P_R e_j \right)U_1^\mu + \textrm{h.c.}\,,
\label{eq:L}
\end{align}
with $P_{L/R}=(1\mp\gamma_5)/2$
in the fermion mass eigenbasis.
Although 
additional vector-like fermions are needed in Eq.~\eqref{eq:L} 
to obtain the ideal flavor structure in the  UV complete model \cite{Cornella:2019hct},
we focus on the $3\times 3$ flavor structures. 
This simplification is valid to consider 
 the EDMs, and we will discuss this point in Sec.~\ref{sec:other}. 
This class of UV complete model is called 4321 gauge-group model \cite{DiLuzio:2017vat,Baker:2019sli}.

We consider the following flavor texture \cite{Bordone:2018nbg,Cornella:2019hct} 
 \begin{align}
 \begin{aligned}
 \beta_L^{ij} &\simeq
 \beta_L^{33}
  \begin{pmatrix}
    0&0&- c_d s_{q_2} s_\chi \left|\frac{V_{td}}{V_{ts}}\right|\\
    0&0&c_d s_{q_2} s_\chi\\
    0&0&c_\chi \end{pmatrix}
   \,, \\
  \beta_R^{ij} &\simeq \beta_L^{33} e^{i \phi_R}
  \begin{pmatrix}
    0&0&0\\
    0&0&0\\
    0&0&1
  \end{pmatrix}\,,
\label{eq:coupling_structure}
\end{aligned}
\end{align}
where $c_d\simeq 0.98$,
corresponding to a case of $s_{l_2}\simeq s_{\tau} \simeq 0$ in the literature. 
Here, $s_i$ and $c_i$ represent flavor rotations $\sin\theta_i$ and $\cos\theta_i$ to bring the SM fermions to their mass eigenbasis.
Note that $|\beta_{L}^{33}|\simeq|\beta_{R}^{33}|$ results from the gauge symmetry in the UV complete model.
In this setup, 
$\phi_R$ is an arbitrary CPV phase; the other CPV phases can be absorbed by a redefinition of $\phi_R$ \cite{Iguro:2018vqb}.
Therefore, the relative phase between $\beta_L$ and $\beta_R$ interactions plays an important role in the CPV observables.
In our analysis, only three parameters are relevant to the phenomenology:  $\beta_L^{33}/m_{U_1}$, $\beta_L^{23}/\beta_L^{33}$\,$(= c_d s_{q_2} s_\chi/c_\chi)$, and the CPV angle $\phi_R$.

\section{EDMs and other observables}
\label{sec:EDM}

In this section, we concisely summarize the phenomenological effects of the $U_1$ LQ.

First, we focus on the LQ contributions to EDMs.
The effective Lagrangian for the EDM ($d_f$) and chromo-EDM interactions ($\tilde{d}_f$) are expressed as 
\begin{align}
\! \!\!\!\!\!\!\!\!{\cal {L}}_{\rm eff} \!= \! -\frac{i}{2}\sum_{f}\left(d_f \ov{f} \sigma_{\mu\nu} \gamma_5 f F^{\mu\nu}
    + g_s\tilde{d}_f \ov{f} \sigma_{\mu\nu} T^a \gamma_5 f G^a_{\mu\nu}\right),\!
\end{align}
with $\sigma_{\mu \nu}= \frac{i}{2}[\gamma_\mu, \gamma_\nu]$.
Based on Refs.~\cite{Queiroz:2014zfa,Kowalska:2018ulj,Altmannshofer:2020ywf},
the $U_1$ LQ contributions to the tau-lepton and bottom-quark (chromo-) EDMs  
are (see Fig.~\ref{fig:dia} right diagram)
\begin{align}
    d_\tau &=-\frac{3e}{8\pi^2} \frac{m_b (\Lambda_{\rm LQ})}{m_{U_1}^2} \textrm{Im}\left[\beta_L^{33} (\beta_R^{33})^*\right]\,,\\
    d_b(\Lambda_{\rm LQ})&= 
 -\frac{5 e}{24 \pi^2} \frac{m_\tau}  {m_{U_1}^2}
     \textrm{Im}\left[\beta_L^{33} (\beta_R^{33})^*\right]\,,\\
    \tilde{d}_b(\Lambda_{\rm LQ})&=-\frac{1 }{8\pi^2} \frac{m_\tau}{m_{U_1}^2} \textrm{Im}\left[\beta_L^{33} (\beta_R^{33})^*\right]\,,
\end{align}
and there is no contribution to the other EDMs at the LQ mass scale, $\mu = \Lambda_{\rm LQ}$.
Note that the Weinberg operator 
would be induced at two-loop level, but it is suppressed by $m_b m_\tau/m_{U_1}^4$, and we discarded it \cite{Dekens:2018bci}.
While the QCD renormalization-group (RG) evolution does not affect $d_\tau$,
we have incorporated the RG evolutions  from $\Lambda_{\rm LQ} $ to $\mu_b (=m_b)$, which
is known to relax the EDM bound \cite{Dekens:2013zca}.
Including the first nontrivial photon-loop effect \cite{Jenkins:2017dyc,Gisbert:2019ftm}, for $\Lambda_{\rm LQ} = 2\,$TeV, we obtain \cite{Degrassi:2005zd,Crivellin:2019qnh,Haisch:2021hcg}
\beq
d_b (\mu_b) &= 0.82\,d_b(\Lambda_{\rm LQ}) + 0.21e\,\tilde{d}_b(\Lambda_{\rm LQ}) \,,\\
\tilde{d}_b (\mu_b) &= 0.08 \frac{d_b(\Lambda_{\rm LQ})}{e}
+ 0.90\,\tilde{d}_b(\Lambda_{\rm LQ}) \,.
\eeq

After integrating out the tau and bottom quark at low energy, 
the electron EDM is induced by the tau and bottom-quark EDMs from QED three-loop radiative corrections \cite{Grozin:2008nw}.
Furthermore, a semi-leptonic $CP$-odd operator, $(\ov{e}i\gamma_5 e)(\ov{p}p + \ov{n}n)$, is also induced from QED two-loop diagrams \cite{Ema:2021jds,Ema:2022pmo}, 
which eventually mimics the electron EDM (called an equivalent electron EDM) in the experiments \cite{Pospelov:2013sca,Kaneta:2023wrl}.
By using a result of the improved analysis for the three-loop calculation \cite{Ema:2022wxd}, 
we obtain
\begin{align}
    d_e^{\rm equiv} &=  \left[4.7\times 10^{-13}+8.8\left(1\pm0.1\right)\times 10^{-12}\right]\, d_b (\mu_b) \nonumber \\
    &\quad +  \left(9.9 \times 10^{-12} + 9.2\times 10^{-14}\right)\, d_\tau \,.
    \label{eq:eEDM}
\end{align}
Here, the first terms in each parenthesis come from the QED three-loop contribution, while the second terms come from 
the semi-leptonic $CP$-odd operator \cite{Ema:2022pmo}.
Note that the latter calculation is a result in the case of the HfF$^{+}$ molecule system \cite{Roussy:2022cmp} (see Refs.~\cite{Pospelov:2013sca,Kaneta:2023wrl} for the other molecules). 
The dominant theoretical uncertainty comes from 
the semi-leptonic $CP$-odd operator
induced by the bottom-quark EDM, which is estimated as 
$10\%$ \cite{Ema:2022pmo}.

By a similar but more involved processes, the nucleon (neutron and proton) EDMs are induced from the bottom-quark EDM and chromo-EDM.
Short-distance contributions come from the light-quark EDM and chromo-EDMs, $d_N^{\rm light}$, and the Weinberg operator, 
 $d_N^{\rm W}$ \cite{Boyd:1990bx,Braaten:1990gq,Chang:1990jv}, while a long-distance contribution arises from a $CP$-odd photon-gluon operator ($GGG\tilde{F}$), $d_N^{\tilde{F}G^3}$ \cite{Ema:2022pmo}. 
For the neutron and proton EDMs,  we numerically obtain
\begin{align}
   d_N&=d_N^{\rm light}+d_N^{\rm W} + d_N^{\tilde{F}G^3} \qquad (\textrm{for~}N=n,\,p)\,,
 \end{align}
 with
\begin{align} 
   d_n^{\rm light}&=  4.0 \times 10^{-7} e\, \tilde{d}_b (\mu_b) + 4.0\times 10^{-8} \, d_b(\mu_b) \,,\\
   d_p^{\rm light}&=  -3.3 \times 10^{-7} e\, \tilde{d}_b (\mu_b) + 9.1\times 10^{-9} \, d_b(\mu_b) \,,\\   
   d_n^{\rm W}& = -5.4\left(1\pm0.5\right)\times 10^{-5} e\, \tilde{d}_b (\mu_b)\,,
   \label{eq:dnW}\\
    d_p^{\rm W}& = 7.7\left(1\pm0.5\right)\times 10^{-5} e\, \tilde{d}_b (\mu_b)\,,
    \label{eq:dpW}\\
    d_N^{\tilde{F}G^3}& \approx 
    7\times 10^{-7}\, d_b (\mu_b)\qquad (\textrm{for~}N=n,\,p)\,.
\end{align}

For $d_N^{\rm light}$, 
the QCD sum-rule estimate is used \cite{Pospelov:2000bw,Pospelov:2005pr,Hisano:2012sc,Fuyuto:2013gla,Kaneta:2023wrl} (where the Peccei-Quinn mechanism is assumed to suppress the $\bar{\theta}$ parameter), whose overall normalization is determined by the lattice result \cite{Alexandrou:2019brg}.
The light-quark EDMs are induced by the bottom-quark EDM \cite{Ema:2022pmo} and chromo-EDM \cite{Ema:2022wxd},
while the  light-quark chromo-EDMs are induced from the bottom-quark chromo-EDM \cite{Haisch:2021hcg}.
For $d_N^{\rm W}$, the QCD sum-rule estimates \cite{Demir:2002gg,Haisch:2019bml,Kaneta:2023wrl} (see also \cite{Yamanaka:2020kjo}) are used.
Note that although all the above terms have $10\%$--$30\%$ theoretical uncertainties, we suppressed them except for the leading one ($d_N^{\rm W}$).
For $d_N^{\tilde{F}G^3}$, the QCD sum-rule technique is also used 
 and the numerics should be understood as an order-of-magnitude estimation \cite{Ema:2022pmo}.

\begin{table}[t]
\begin{center}
  \begin{tabularx}{0.48 \textwidth}{cYl} 
   ~~EDM [$e\,$cm] & $90\%$ CL limit  & Future sensitivity~~ \\ \hline
  ~~$|d_e|$ & $\le 4.1\times 10^{-30}$ \cite{Roussy:2022cmp} & $ \mathcal{O}(10^{-31})$ \cite{Alarcon:2022ero}\\ 
  ~~$|d_n|$ & $\le 1.8\times 10^{-26}$ \cite{Abel:2020pzs} & 
 $ \mathcal{O}(10^{-27})$ \cite{n2EDM:2021yah,Martin:2020lbx,Ito:2017ywc,Wurm:2019yfj}\\
 & & 
  $ \mathcal{O}(10^{-28})$ \cite{nEDM:2019qgk}\\ 
  ~~$|d_p|$ & $\le 2.1\times 10^{-25}$ \cite{Sahoo:2016zvr} & $ \mathcal{O}(10^{-29})$ \cite{CPEDM:2019nwp,Alexander:2022rmq}\\ \hline
   \end{tabularx}
  \caption{The current $90\%$ confidence level (CL) upper limits and future prospects for electron, neutron, and proton EDMs.
   }
  \label{Tab:EDM}
\end{center}   
\vspace{-.45cm}
\end{table}

It is found that the overwhelmingly dominant contribution to the nucleon EDMs comes from the Weinberg operator. 
Also, the theoretical uncertainty is dominated by the Weinberg operator, which is estimated as $50\%$ \cite{Haisch:2019bml}.
Although the accuracy of the lattice calculations is currently not competitive \cite{Bhattacharya:2016rrc,Yoon:2017tag,Mereghetti:2018sxs,Yoon:2020soi,Todaro:2021cal,Bhattacharya:2022whc,Bhattacharya:2023xov},
they will provide complementary inputs in the future.
We emphasize that the predicted neutron and proton EDMs must be the same size with opposite signs \cite{Haisch:2019bml}.

The current upper bounds and the future prospects for the electron, neutron, and proton EDMs are summarized in Table~\ref{Tab:EDM}.

\subsection{\texorpdfstring{\boldmath{$R_{D^{(*)}}$}}{RD(*)}}

The $U_1$ LQ can naturally explain the $R_{D^{(*)}}$ anomalies.
After integrating out the LQ and the weak bosons, the  effective Lagrangian is given by
\begin{align}
 \label{eq:Hamiltonian}
 {\cal {L}}_{\rm{eff}}= -2 \sqrt2 G_FV_{cb}\left[ \left(1+C_{V_L}\right)O_{V_L}+C_{S_
R}O_{S_R}\right]\,,
\end{align}
with
$O_{V_L} = (\ov{c} \gamma^\mu P_L b)(\ov{\tau} \gamma_\mu P_L \nu_{\tau})$, $ O_{S_R}  = (\ov{c}  P_R b)(\ov{\tau} P_L \nu_{\tau})$,
and the Wilson coefficients (WCs) at $\mu=\mu_b$ are
\begin{align}
    C_{V_L} (\mu_b)&= \frac{\eta_{V_L}}{2\sqrt{2} G_F V_{cb}}
    \frac{\beta_{L}^{23}(\beta_{L}^{33})^{*}}{m_{U_1}^2}\,,\\
    C_{S_R} (\mu_b)&=- \frac{\eta_{S_R}}{\sqrt{2} G_F V_{cb}} \frac{\beta_{L}^{23}(\beta_{R}^{33})^{*}}{m_{U_1}^2}\,,
\end{align}
where $\eta_{V_L}$ and $\eta_{S_R}$ are coefficients of the QCD corrections \cite{Aebischer:2017gaw,Gonzalez-Alonso:2017iyc,Aebischer:2018acj}. For $\Lambda_{\rm LQ} \simeq 2$--$4$\,TeV, 
$\eta_{V_L}\simeq 1.1$ and $\eta_{S_R}\simeq 2.0$ \cite{Iguro:2024hyk}.
Furthermore, 
assuming 
the simplified flavor texture in Eq.~\eqref{eq:coupling_structure},
these two WCs can be correlated with being 
\begin{align}
    C_{S_R}(\mu_b) \simeq - 3.6 \, e^{-i\phi_R} C_{V_L}(\mu_b)\,.
\end{align}

\begin{figure}[t]
    \centering
 \includegraphics[width=0.45\textwidth]{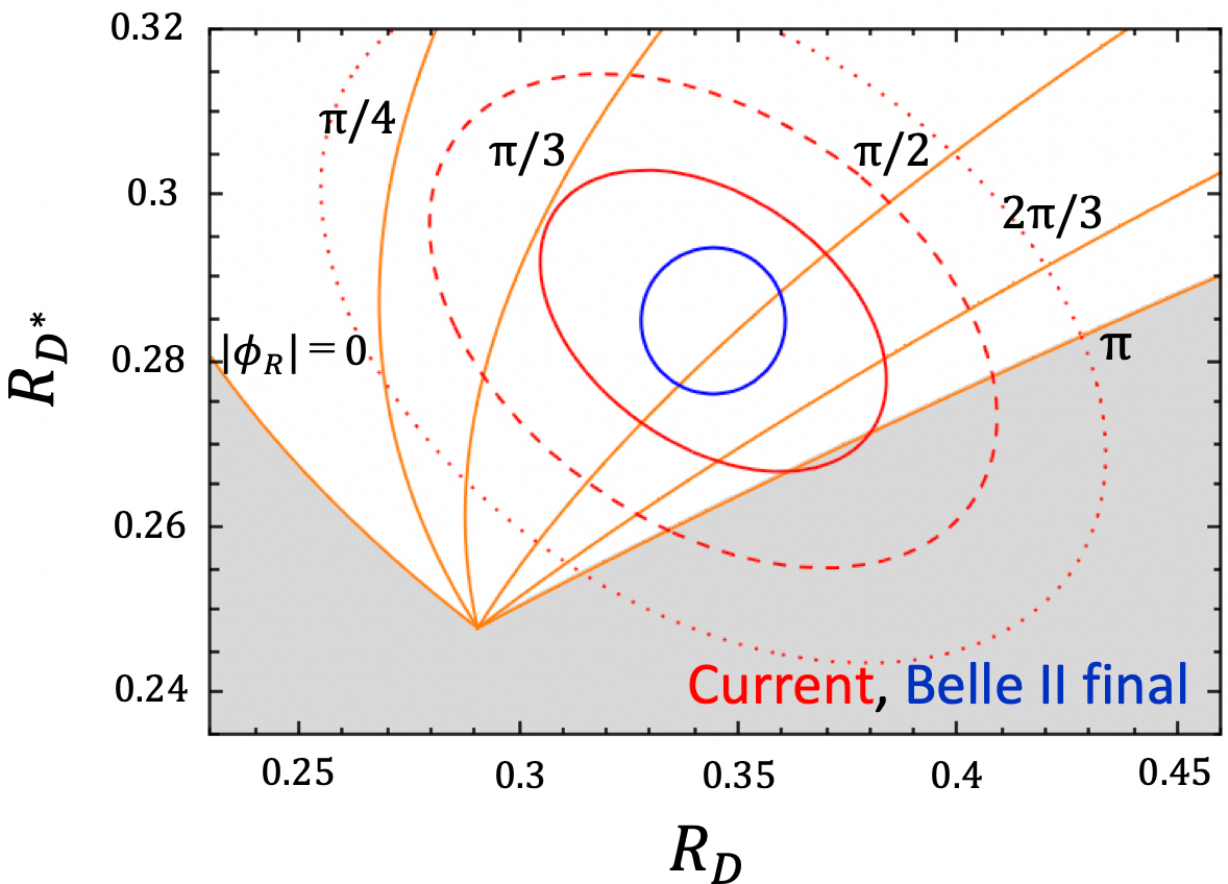}
     \vspace{-.2cm}
\caption{The orange contour represents 
the CPV phase $|\phi_R|$ 
on the plane of $R_{D}$--$R_{D^{*}}$ in the simplified $U_1$ scenario.
The red solid, dashed, and dotted contours correspond to $1,\,2,\,3\sigma$ of the experimental world average \cite{HFLAV2024winter}.
The blue circle denotes a sensitivity projection of the Belle II experiment \cite{Belle-II:2018jsg} assuming the current central values. 
The gray-shaded region is out of the model prediction. 
}
\label{fig:RDvsU2}
   \vspace{-.4cm}
\end{figure}
By using the numerical formulae for $R_{D^{(*)}}$ in Ref.~\cite{Iguro:2024hyk}, based on the heavy quark effective theory  form factors \cite{Iguro:2020cpg}, 
we show a correlation between $R_{D^{(*)}}$ and the CPV phase $\phi_R$ 
in Fig.~\ref{fig:RDvsU2}.
Since $R_{D^{(*)}}$ are the $CP$-conserving observables, they depend on only $\cos\phi_R$ and are invariant under $\phi_R \leftrightarrow -\phi_R$.
The orange contour denotes the values of $|\phi_R|$ with varying $\beta_L^{33}/m_{U_1}$. We use $\beta_L^{23}/\beta_L^{33} = \lambda \simeq 0.225$ as a typical reference value \cite{Cornella:2019hct}. 
The gray-shaded region cannot be predicted within the simplified $U_1$ LQ model.
It is found that large $\phi_R$ ($\pi/3 <|\phi_R|$) is favored to accommodate the $R_{D^{(*)}}$ anomaly, 
while a $CP$-conserving scenario of $\phi_R=0$ can be excluded by the current data.
One should note that 
the $U_1$ LQ model also leads to deviations from the SM predictions in other $b\to c\tau \ov\nu$ observables, $\tau$ polarization asymmetry and the LFU violation in $\Lambda_b\to\Lambda_c l \ov\nu$, which will be shown in Appendix~\ref{sec:Appendix_other_bctaunu}.

\begin{figure*}[t]
\includegraphics[width=0.45\textwidth]{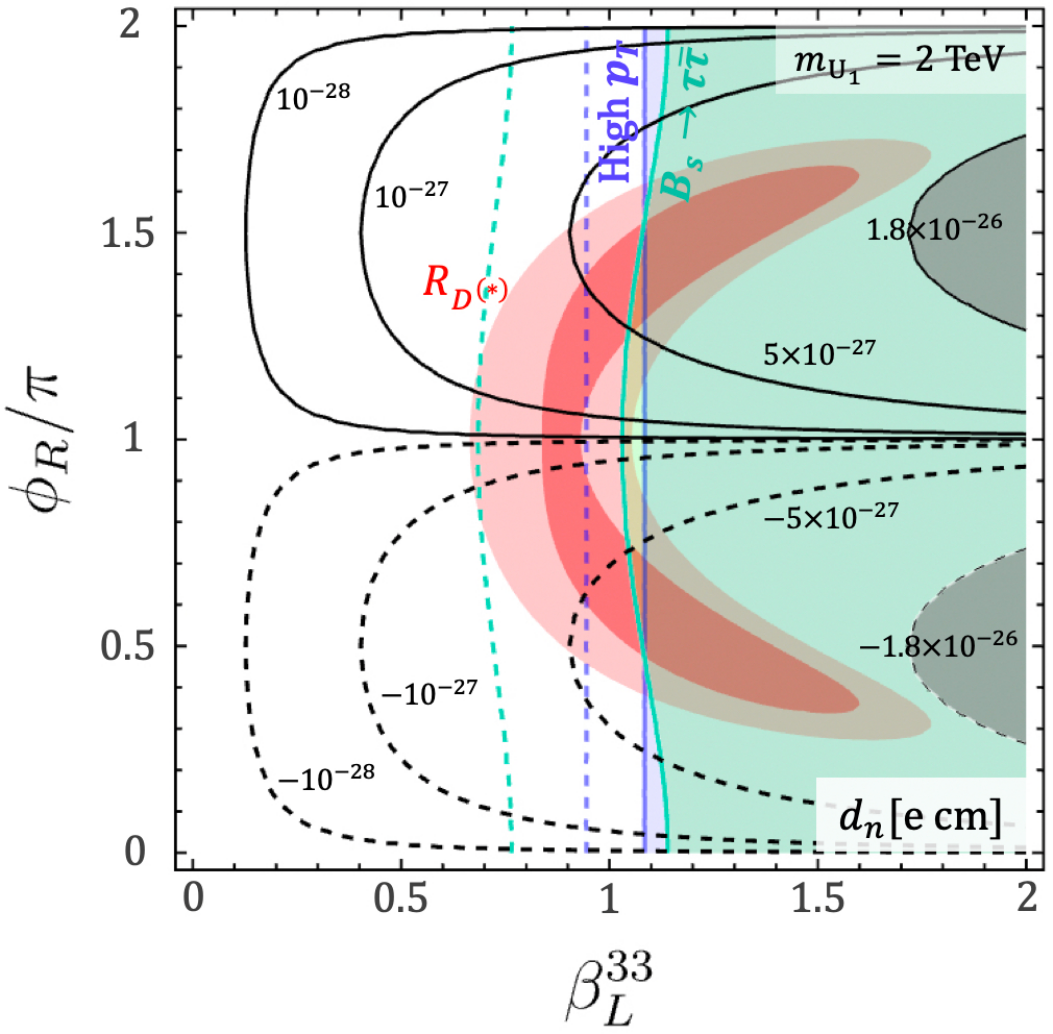}
\qquad 
\includegraphics[width=0.45\textwidth]{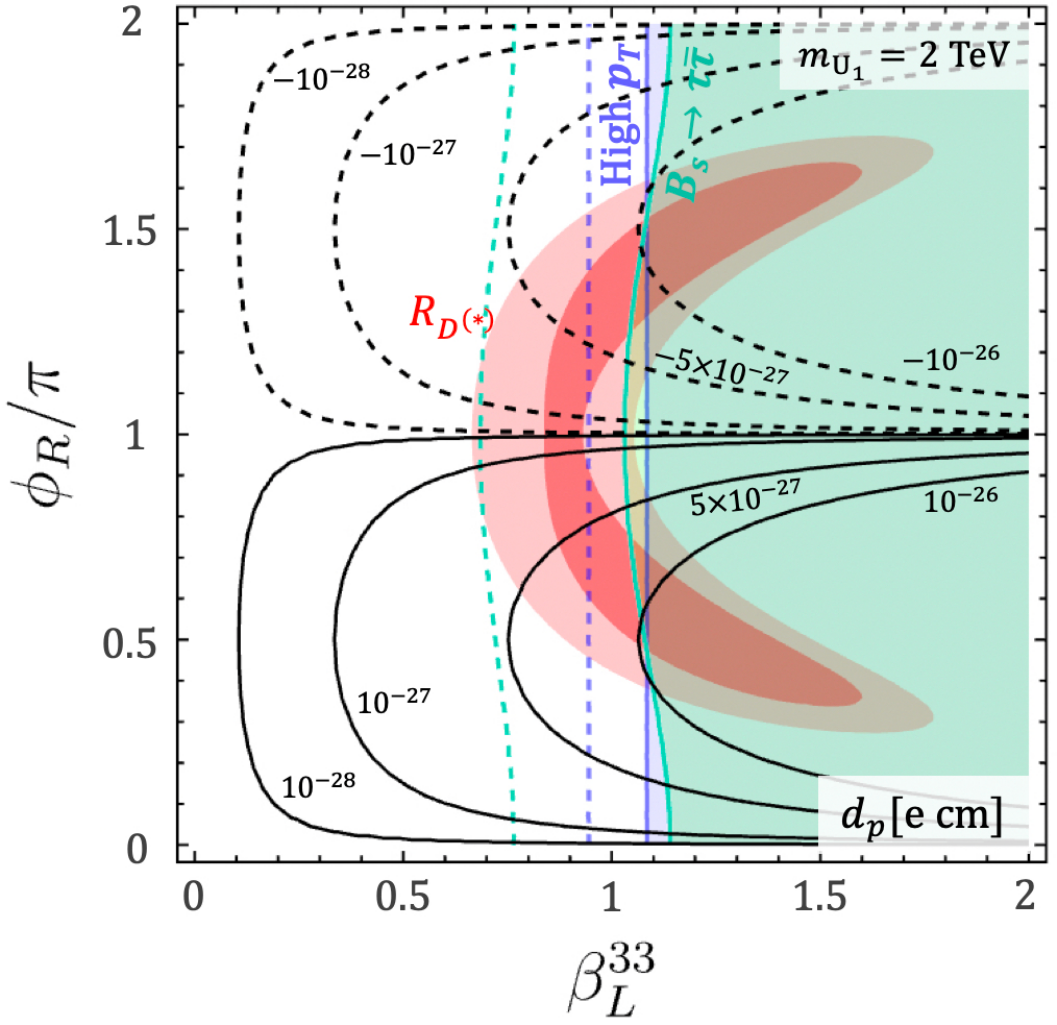}
\caption{\label{ref:param}
The predicted neutron and proton EDMs are shown by the black contours in the left and right panels, respectively, where
the solid (dashed) lines represent positive (negative) EDMs, in the $U_1$ scenario. The gray-shaded region is excluded by the current upper bound on the neutron EDM \cite{Abel:2020pzs}.
The red (light red) region can explain the \rdst\,anomaly at 1$\sigma$ (2$\sigma$) level.
The blue and green regions are excluded by the high-$p_T$ bound and
$B_s\to\tau^+\tau^-$, respectively.
The estimated sensitivities based on upcoming Run\,3 data are shown by the dashed blue and green lines.
We set $m_{U_1}=2\,$TeV and $\beta_L^{23}/\beta_L^{33} = \lambda$.
}
\end{figure*}

\subsection{\texorpdfstring{\boldmath{$B_s\to\tau^+\tau^-$}}{Bs to tau tau}}

Within the SM, $B_s\to\tau^+\tau^-$ is suppressed by the one-loop factor and also the chirality factor, $m_\tau^2/m_{B_s}^2$.
On the other hand, the $U_1$ LQ contributions are induced at the tree level 
and further the chirality suppression can be avoided.
Therefore, $B_s\to\tau^+\tau^-$ is significantly affected by the LQ.
Currently, the LHCb with Run\,1 data sets the upper limit on the branching ratio 
 at 95\,$\%$ CL as \cite{LHCb:2017myy} 
\begin{align}
\mathcal{B}(B_s\to\tau^+\tau^-)\le6.8\times10^{-3}\,.
\end{align}
The future prospect of the LHCb Run\,3 has been estimated to improve the sensitivity by a factor of five \cite{LHCb:2018roe}. 
The $U_1$ LQ contribution to $B_s\to\tau^+\tau^-$ including the QCD corrections is give by \cite{Cornella:2019hct}
\begin{align}
    &\frac{
    \mathcal{B}(B_s\to\tau^+\tau^-)}
    {\mathcal{B}(B_s\to\tau^+\tau^-)^{\rm{SM}}}
   \nonumber \\
   &\simeq
     \biggl|1+\frac{\pi }{\sqrt{2} \alpha G_F V_{tb}V_{ts}^*m_{U_1}^2}\beta_L^{23}\left(-0.26\beta_L^{33}+1.8\beta_R^{33}\right)^*\biggl|^2\nonumber\\
    &\quad +\left(1-\frac{4m_\tau^2}{m_{B_s}^2}\right)\biggl|\frac{1.8 \pi}{\sqrt{2} \alpha G_F V_{tb}V_{ts}^*m_{U_1}^2}\beta_L^{23}(\beta_R^{33})^{*}\biggl|^2\,.
\end{align}
It is noted that the effect from the CPV phase $\phi_R$ is mild due to the smallness of the SM contribution.

\subsection{LHC high-\texorpdfstring{\boldmath{$p_T$}}{pT} bound}
We employed a public tool \texttt{HighPT} \cite{Allwicher:2022mcg} to derive the collider constraint from $pp \to \tau^+ \tau^-$ and $pp \to \tau \nu$ data.
Currently, the dominant constraint comes from the high-$p_T$ di-$\tau$ search from the ATLAS collaboration \cite{ATLAS:2020zms} (see also Refs.~\cite{Baker:2019sli,Bhaskar:2021pml} for the relevant study).
At the CMS, an excess has been found in the high-$p_T$ tail region \cite{CMS:2022zks,CMS:2022goy}.
However, the ATLAS does not find excess in the region.\footnote{More detailed experimental comparisons and/or statistics are necessary to conclude the difference between the CMS and ATLAS results \cite{CMSLP2023}.}
On the other hand, the constraint from high-$p_T$ mono-$\tau$ search is currently less constraining \cite{Greljo:2018tzh,Iguro:2020keo}.
However, it has been pointed out that requiring an additional $b$-tagged jet can improve the sensitivity so that this channel is competitive with the di-$\tau$ channel \cite{Marzocca:2020ueu,Endo:2021lhi}.

\begin{figure*}[t]
\includegraphics[width=0.45\textwidth]{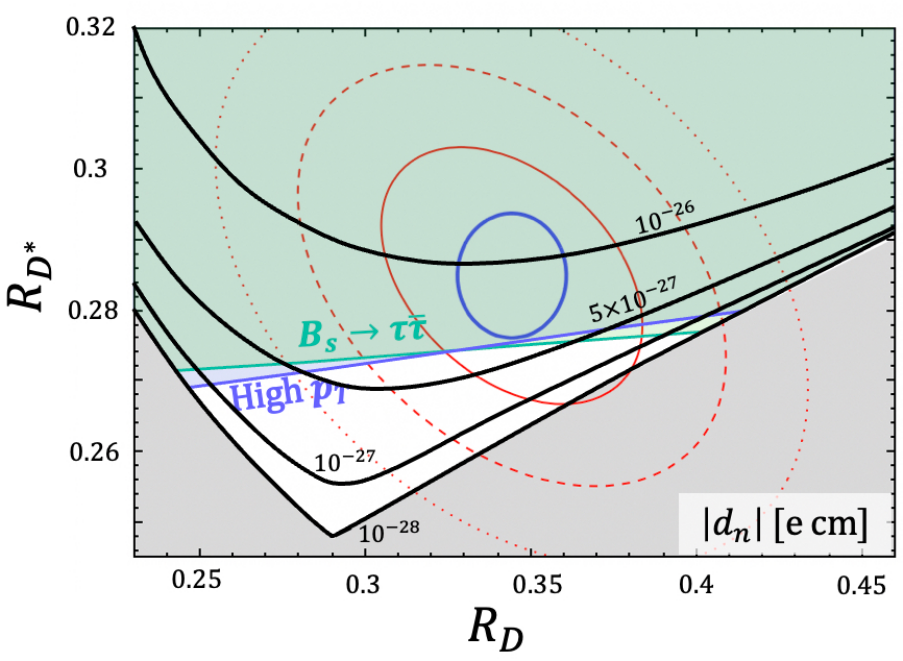}
\qquad 
\includegraphics[width=0.45\textwidth]{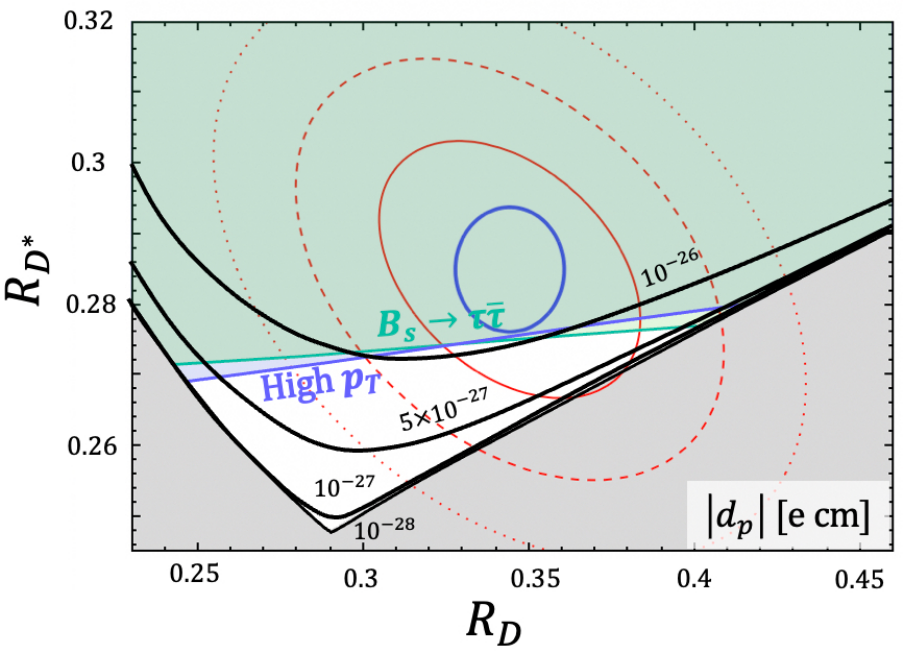}
\caption{\label{fig:RD_EDM}
The absolute values of the predicted neutron and proton EDMs 
are shown by
the black contours in the left and right panels, respectively, in the $U_1$ scenario.
Constraints from the high-$p_T$ search and $B_s\to\tau^+\tau^-$ are represented by the blue and green regions, respectively.
The gray-shaded regions are out of the model prediction. 
We set $m_{U_1}=2\,$TeV and $\beta_L^{23}/\beta_L^{33} = \lambda$.
}
\end{figure*}

\subsection{Comment on other constraints}
\label{sec:other}

It is known that although loop-induced LQ contributions to \bbms~give a severe constraint, 
once additional vector-like fermions are introduced in the UV complete model 
the constraint can be naturally avoided thanks to the GIM-like mechanism \cite{Calibbi:2017qbu,DiLuzio:2018zxy,Fuentes-Martin:2020hvc,Marzocca:2018wcf,Crivellin:2018yvo,Cornella:2019hct,Iguro:2022ozl}.
We emphasize that the vector-like fermions do not mix the SM right-handed fermions in the UV complete model, and the EDMs are not induced from the vector-like fermion loops \cite{Cornella:2019hct}.
Therefore, the EDMs provide a unique prediction of the model.

The similar sensitivity to $B_s\to\tau^+\tau^-$ could be obtained from the measurement of $B \to K \tau^+\tau^-$ at the Belle II \cite{Capdevila:2017iqn}, while we omitted it since the current bound is much weaker.
Although $B^-\to\tau\ov\nu$ is also modified in the simplified flavor texture,
a moderate $\beta_L^{13}$ suppresses the constraint \cite{Cornella:2019hct}.

Additionally, we would like to comment on the possibility of further contributions to the EDMs from other scalar particles, which are required for the gauge symmetry breaking. 
In the 4321 model, in addition to the $U_1$ LQ, the $R_2$ scalar LQ is also introduced (it is called $H_{15}$ field in Ref.~\cite{Cornella:2019hct}). 
This $R_2$ LQ could produce two large EDMs: (1) top-quark and (2) tau-lepton EDMs. We estimated both contributions and found that (1) predicted top-quark EDM is two orders of magnitude smaller than the current experimental bound $|d_t| <5 \times 10^{-20}\,e\,$cm~\cite{Cirigliano:2016njn}, and (2) the predicted tau-lepton EDM is two orders of magnitude larger by a factor of $m_t/m_c$ than Eq.~\eqref{eq:dtauR2}. The latter induces the electron EDM at $\mathcal{O}(10^{-30})\,e\,$cm, which
can be probed by future experiments, see Table~\ref{Tab:EDM}.
Note that this $R_2$ LQ in the 4321 model can not account for the $R_{D^{(\ast)}}$ anomaly.

\section{Result of the \texorpdfstring{\bm{$U_1$}}{U1} LQ scenario}
\label{sec:resultU1}
In Figs.~\ref{ref:param}
and \ref{fig:RD_EDM},
we show the correlations between the predicted nucleon (neutron and proton) EDMs and the $R_{D^{(*)}}$ anomaly in the $U_1$ LQ model.
Here, $m_{U_1}=2\,$TeV and $\beta_L^{23}/\beta_L^{33} = \lambda$ are taken as reference values. 
Black contours in Fig.~\ref{ref:param} indicate the neutron and proton EDMs in the left and right panels, respectively, where
the solid (dashed) lines represent positive (negative) EDMs.
We used the central values of  Eqs.~\eqref{eq:dnW} and \eqref{eq:dpW} for the estimates of the nucleon EDMs.
The blue and green regions are excluded by the high-$p_T$ bound and
$B_s\to\tau^+\tau^-$, respectively.
The estimated sensitivities based on upcoming Run\,3 data are shown by the dashed blue and green lines in Fig.~\ref{ref:param}.
It is noted that $B_s\to \tau^+\tau^-$ at Run\,3 will be able to cover most of the preferred parameter region at the $2\sigma$ level.
We also show the correlations on the $R_D$--$R_{D^*}$ plane in Fig.~\ref{fig:RD_EDM}.

These figures show that 
some of the preferred areas are already excluded by both the high-$p_T$ bound and
$B_s\to\tau^+\tau^-$.
In the allowed regions, the predicted magnitudes of the nucleon EDMs are
$|d_n|< 7\times 10^{-27}\,e\,$cm and $|d_p| < 1\times 10^{-26}\,e\,$cm.
Very excitingly, in the near future, several experiments will probe the neutron EDM at $\mathcal{O}(10^{-27})\,e\,$cm  precision \cite{n2EDM:2021yah,Martin:2020lbx,Ito:2017ywc,Wurm:2019yfj}, and eventually $\mathcal{O}(10^{-28})\,e\,$cm \cite{nEDM:2019qgk}.
Furthermore, two experiments are proposed that the proton EDM will be proved at $\mathcal{O}(10^{-29})\,e\,$cm precision \cite{CPEDM:2019nwp,Alexander:2022rmq}.
Therefore, we conclude that nucleon EDMs with their opposite signs and $B_s\to \tau^+\tau^-$ will be a smoking-gun signal of the $U_1$ LQ model.

On the other hand,
the induced electron EDM from Eq.~\eqref{eq:eEDM} is $|d_e| < 10^{-32}\,e\,$cm, which 
is a few orders away from the future prospect, but the suppressed electron EDM is also a unique prediction of this model.

\section{Scalar \texorpdfstring{\boldmath{$R_2$}}{R2} LQ scenario}
\label{sec:R2}

In this section, we perform the same analysis as the main text for the $R_2$ LQ scenario. The gauge charge of the scalar LQ $R_2$
is $(\bm{3},\bm{2},7/6)$ and the $SU(2)_L$-doublet $(R_2^{\frac{5}{3}},\,R_2^{\frac{2}{3}})$ has a common mass $m_{R_2}$.
The fermion interactions are described by 
\beq
\mathcal{L}=
 - y_L^{ij} \ov{u}_{i} R_2^T \epsilon P_L L_{j} 
+ y_R^{ij} \ov{Q}_{i} R_2 P_R e_{j} +\mathrm{h.c.}\,,
\eeq
with $\epsilon_{12}=1$.

\begin{figure*}[t]
\includegraphics[width=0.45\textwidth]{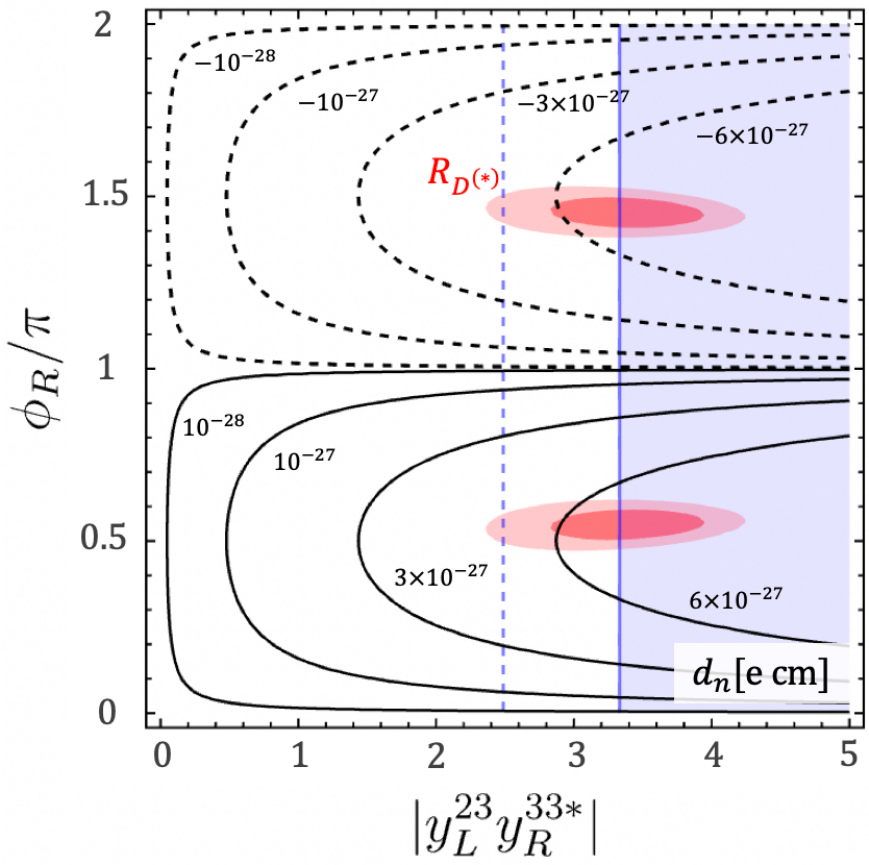}
\qquad 
\includegraphics[width=0.45\textwidth]{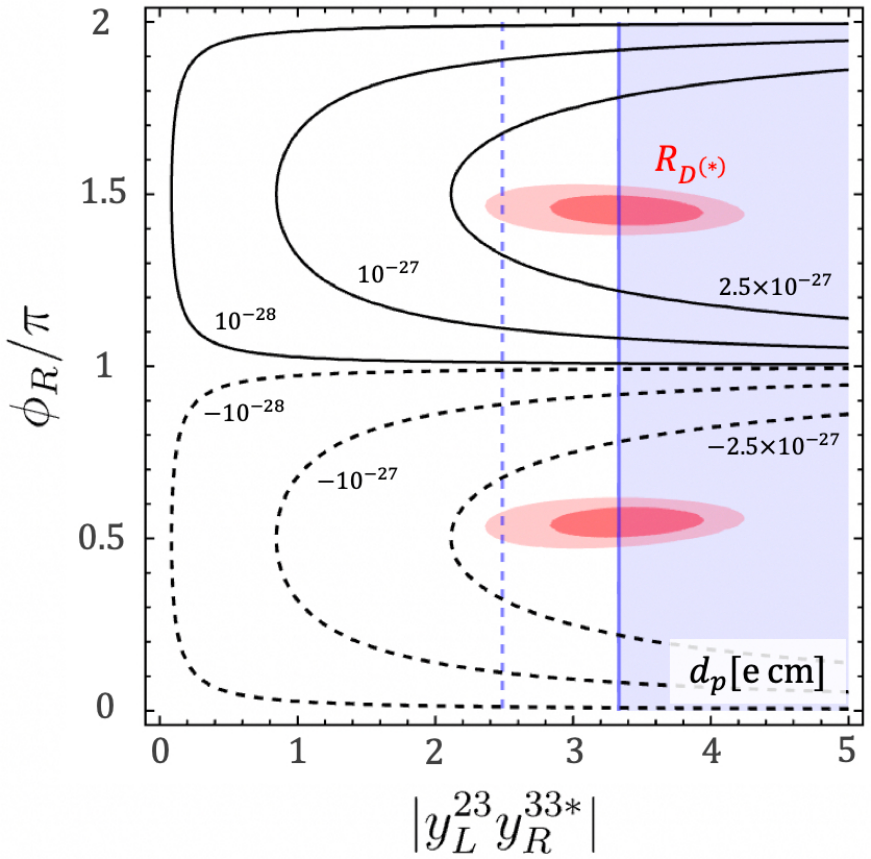}
\caption{\label{ref:param_R2}
The predicted neutron and proton EDMs are shown in the $R_2$ scenario.
The red (light red) region can explain the \rdst\,anomaly at 1$\sigma$ (2$\sigma$) level.
The blue and green regions are excluded by the high-$p_T$ bound.
The estimated sensitivities based on upcoming Run\,3 data are shown by the dashed blue lines.
We set $m_{R_2}=2\,$TeV and $|y_L^{23}|/|y_R^{33}| = 0.7$.}
\vspace{.3cm}
%
\includegraphics[width=0.45\textwidth]{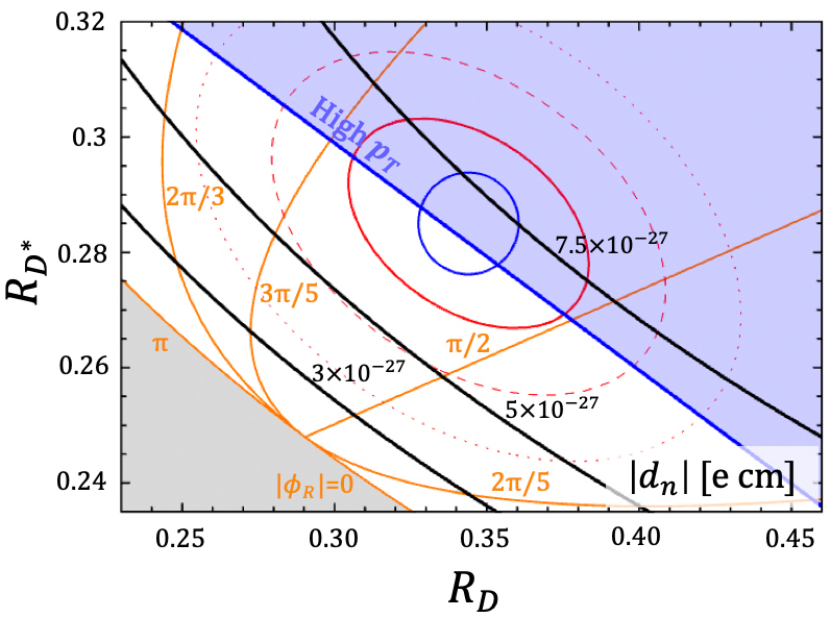}
\qquad 
\includegraphics[width=0.45\textwidth]{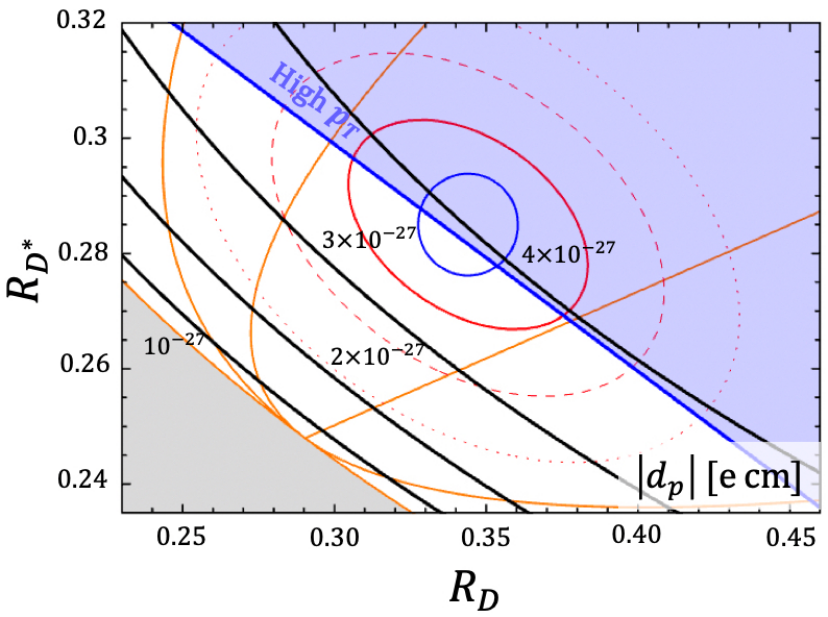}
\caption{\label{fig:RD_EDM_R2}
The absolute values of the predicted neutron and proton EDMs 
are shown by the black contours in the left and right panels, respectively, in the $R_2$ scenario.
The orange contour represents the CPV phase $|\phi_R|$.
The blue regions are excluded by the high-$p_T$ bound.
We set $m_{R_2}=2\,$TeV and $|y_L^{23}|/|y_R^{33}| = 0.7$.
}
    \vspace{.3cm}
%
%
\includegraphics[width=0.45\textwidth]{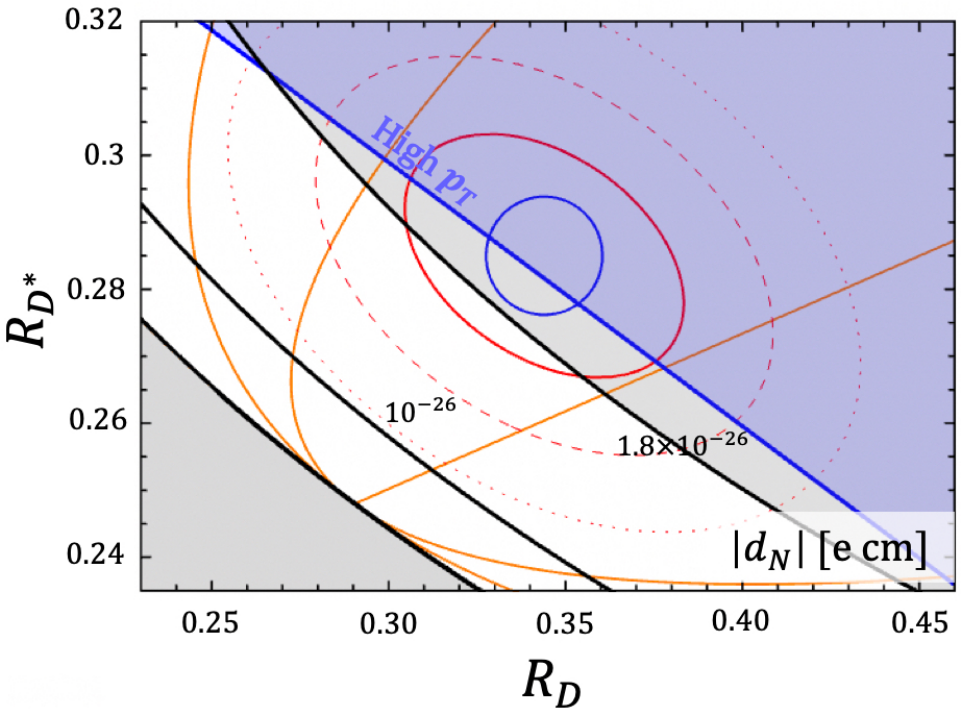}
\qquad 
\caption{\label{fig:RD_EDM_dN}
Same as the Fig.~\ref{fig:RD_EDM_R2}, but the
nucleon EDM is evaluated by the lattice charm tensor charge.
The gray-shaded region in upper right is excluded by the current upper bound on the neutron EDM \cite{Abel:2020pzs}.
}
\end{figure*}

The tree-level $R_2^{\frac{2}{3}}$ exchange contributes to the $b \to c \tau \ov\nu_\tau$ process 
with the WCs of \cite{Iguro:2024hyk}
\beq
C_{S_L} (\Lambda_{\rm LQ}) = 4 C_T(\Lambda_{\rm LQ}) =\frac{1}{4\sqrt{2}G_F V_{cb}}\frac{y_L^{23} (y_R^{33})^\ast}{m_{R_2}^2}\,,
\eeq
where the normalization of the effective Lagrangian is the same as that in the main text  with  
\beq
 O_{S_L}  &= (\ov{c}  P_L b)(\ov{\tau} P_L \nu_{\tau})\,,\\
  O_{T} &= (\ov{c} \sigma^{\mu\nu} P_L b)(\ov{\tau} \sigma_{\mu\nu} P_L \nu_{\tau})\,.
\eeq
Through the RG evolution, the WCs at the low energy are obtained as \cite{Iguro:2024hyk}
\beq
C_{S_L} (\mu_b) &= 1.91\, C_{S_L} (\Lambda_{\rm LQ}) - 0.38 \, C_T(\Lambda_{\rm LQ})\,,\\
C_{T} (\mu_b) &=0.89  \,C_T(\Lambda_{\rm LQ})\,,
\eeq
for $\Lambda_{\rm LQ} \simeq 2\,$TeV.

Next, we consider the EDM part. 
Although the $R_2^{\frac{2}{3}}$ component does not contribute to any EDMs,
it is known that 
the $R_2^{\frac{5}{3}}$ LQ affects the tau-lepton and charm-quark (chromo-) EDMs~\cite{Fuyuto:2018scmz,Dekens:2018bci,Babu:2020hun}, 
\beq
d_\tau & = - \frac{e} {32\pi^2} \frac{m_c(\Lambda_{\rm LQ})}{m_{R_2}^2}
\textrm{Im}\left[y_L^{23}V_{cb}^{\ast} (y_R^{33})^*\right] \nonumber \\
&\quad \times \left( 1 + 4 \ln \frac{m_c^2}{m_{R_2}^2} \right)\,,\label{eq:dtauR2}\\
d_c (\Lambda_{\rm LQ}) & = 
\frac{e}{32\pi^2} \frac{m_\tau}{m_{R_2}^2}\textrm{Im}\left[y_L^{23}V_{cb}^{\ast} (y_R^{33})^*\right]
\nonumber \\
&\quad \times \left( \frac{4}{3} +  2\ln \frac{m_\tau^2}{m_{R_2}^2} \right)\,,\\
\tilde{d}_c (\Lambda_{\rm LQ}) & = 
\frac{1}{32\pi^2} \frac{m_\tau}{m_{R_2}^2}\textrm{Im}\left[y_L^{23}V_{cb}^{\ast} (y_R^{33})^*\right]\,.
\eeq
Note that unlike the vector $U_1$ LQ scenario, 
the EDMs receive large logarithmic corrections. 
They are interpreted as the operator mixing  between the scalar-type semi-leptonic operator and dipole operator
in the RG evolution \cite{Hisano:2012cc,Dekens:2018bci}.
At the low energy $\mu_c(=m_c)$, the charm-quark (chromo-) EDMs receive the following RG effects including the first nontrivial QED effect,
\beq
d_c (\mu_c) &= 0.76\,d_c(\Lambda_{\rm LQ}) - 0.58e\,\tilde{d}_c(\Lambda_{\rm LQ}) \,,\\
\tilde{d}_c (\mu_c) &= -0.05 \frac{d_c(\Lambda_{\rm LQ})}{e}
+ 0.87\,\tilde{d}_c(\Lambda_{\rm LQ}) \,,
\eeq
with $\Lambda_{\rm LQ} = 2\,$TeV.

After integrating out the tau and the charm quark, the equivalent electron EDM is induced as 
\begin{align}
    d_e^{\rm equiv} &=  \left[-1.2\times 10^{-11}
    -5.4\left(1\pm0.1\right)\times 10^{-10}\right]\, d_c (\mu_c) \nonumber \\
    &\quad +  \left(9.9 \times 10^{-12} + 9.2\times 10^{-14}\right)\, d_\tau \,.
    \label{eq:eEDM_R2}
\end{align}
Similar to the $U_1$ LQ case,
the nucleon EDMs 
 $d_N=d_N^{\rm light}+d_N^{\rm W} + d_N^{\tilde{F}G^3}$ for $N=n, p$ 
are radiatively induced as 
\begin{align}
   d_n^{\rm light}&=  -2.0 \times 10^{-6} e\, \tilde{d}_c (\mu_c) + 9.2\times 10^{-7} \, d_c(\mu_c) \,,\\
   d_p^{\rm light}&=  -4.8 \times 10^{-7} e\, \tilde{d}_c (\mu_c) + 2.2\times 10^{-7} \, d_c(\mu_c) \,,\\   
   d_n^{\rm W}& = -5.5\left(1\pm0.5\right)\times 10^{-4} e\, \tilde{d}_c (\mu_c)\,,
   \label{eq:dnW_c}\\
    d_p^{\rm W}& = 7.9\left(1\pm0.5\right)\times 10^{-4} e\, \tilde{d}_c (\mu_c)\,,
    \label{eq:dpW_c}\\
    d_N^{\tilde{F}G^3}& \approx 
    3\times 10^{-5}\, d_c (\mu_c)\qquad (\textrm{for~}N=n,\,p)\,.
    \label{eq:dN_charm_nonp}
\end{align}

It must be emphasized that 
the charm quark should not be integrated out for the nucleon EDM evaluations if possible, 
because charm-quark mass is close to the hadronic scale.
It is known that one can investigate the non-perturbative QCD contribution from the charm-quark EDM by using the charm tensor charge $g_T^c$ with $d_N \supset g_T^c d_c$. 
The latest lattice result is $g_T^c =  (-2.4 \pm 1.6)\times 10^{-4}$, where the matching scale is $\mu=2\,$GeV~\cite{Alexandrou:2019brg}.
This has still large uncertainty. Furthermore, 
the value is an order of magnitude larger than the QCD sum-rule result \cite{Ema:2022pmo} [$d_N^{\tilde{F}G^3}$ in Eq.~\eqref{eq:dN_charm_nonp}] with opposite sign.
Therefore, we investigate two different evaluations:
$d_N=d_N^{\rm light}+d_N^{\rm W} + d_N^{\tilde{F}G^3}$ and $d_N = g_T^c d_c$.
References~\cite{Gisbert:2019ftm,Babu:2020hun} investigated the former contribution, while 
Refs.~\cite{Dekens:2018bci,Becirevic:2022tsj,Kirk:2023fin} did the latter one, and none of the literature compares these two contributions.

Similar to the $U_1$ LQ case, 
only a single CPV phase, the relative phase between $y_L^{23}$ and $y_R^{33}$, is relevant.
In this section, we set
$y_L^{23} = |y_L^{23}|$ and  $y_R^{33} = |y_R^{33}|\exp(i \phi_R)$.
Then, the free parameters are only three:
$|y_R^{33}|/m_{R_2}$, $|y_L^{23}|/|y_R^{33}|$, and the CPV angle $\phi_R$.

We also evaluate the high-$p_T$ bound by using  \texttt{HighPT}. 
Note that both components $(R_2^{\frac{5}{3}},\,R_2^{\frac{2}{3}})$ contribute to the bound.

In Figs.~\ref{ref:param_R2} and \ref{fig:RD_EDM_R2}, we show the correlations between the nucleon EDMs and the \RDs\,anomaly 
in the $R_2$ LQ model parameter space and the $R_D$--$R_{D^\ast}$ plane, respectively. 
In both plots, we take
$m_{R_2} = 2\,$TeV and $|y_L^{23}|/|y_R^{33}| = 0.7$. The blue shaded regions are excluded by the  high-$p_T$ analysis. 
We found that the value $|y_L^{23}|/|y_R^{33}| \approx  0.7$  can alleviate this bound, and 
taking 
$|y_L^{23}|/|y_R^{33}| \gg  0.7 $ and $\ll 0.7$ are incompatible to the \RDs\, anomaly at the $1\sigma$ level. 
In Fig.~\ref{ref:param_R2}, the dashed blue lines represent the estimated Run\,3 sensitivity of the high-$p_T$ search.
We also show the CPV phase $|\phi_R|$ by the orange contour in Fig.~\ref{fig:RD_EDM_R2}.
It is clearly shown that the \RDs\,anomaly predicts the large CPV phase $\phi_R$.
In both figures, we show the neutron and proton EDMs calculated by $d_N=d_N^{\rm light}+d_N^{\rm W} + d_N^{\tilde{F}G^3}$. 
In the allowed regions, 
the predicted magnitudes of the nucleon EDMs are
$|d_n|< 7 \times 10^{-27}\,e\,$cm and $|d_p| < 4\times 10^{-27}\,e\,$cm.
It is worth noting that the EDM predictions in  Fig.~\ref{fig:RD_EDM_R2} are insensitive to the ratio of $|y_L^{23}|/|y_R^{33}|$.

On the other hand, in Fig.~\ref{fig:RD_EDM_dN} we plot the nucleon EDM calculated by the charm tensor charge $d_N = g_T^c d_c$.
We found that this evaluation is a factor three to five larger than Fig.~\ref{fig:RD_EDM_R2}.
Also, the $R_{D^{(*)}}$ preferred region is mostly excluded by the current upper bound on $d_n$ \cite{Abel:2020pzs}. 
Note that both estimations ($d_N^{\rm W}$ and $g_T^c d_c$) contain $\mathcal{O}(50)\%$ theoretical uncertainties, which are not included in our analysis.  
Nevertheless, 
we conclude that the predicted nucleon EDMs are  large enough to be observed in the next generation experiments.

Note that there is no $R_2$ LQ contributions to $B_s \to \tau^+\tau^-$, unlike the $U_1$ LQ scenario. 
On the other hand, the $R_2$ LQ contributes to $Z \to \tau^+ \tau^-$ at the one-loop level \cite{Arnan:2019olv,Crivellin:2020mjs}, which has a comparable sensitivity with the high-$p_T$ bound 
\cite{Kirk:2023fin}.

Moreover, We found that the induced electron EDM is $|d_e| < 4\times 10^{-32}\,e\,$cm and it is difficult to probe it by the proposed future experiments. 

For the completeness, we show 
$\tau$ polarization asymmetry and the LFU violation in $\Lambda_b\to\Lambda_c l \ov\nu$ in Appendix~\ref{sec:Appendix_other_bctaunu}.


\section{Summary and Discussion}
\label{sec:summary}

In this paper,
we established a robust bridge between the electric dipole moments and the flavor anomaly in 
$\Bb \to D^{(\ast)} l \overline{\nu}$
through the $SU(2)_L$-singlet vector LQ
coupled primarily to third-generation fermions
as well as the $SU(2)_L$-doublet scalar LQ scenarios. 
In these LQ interactions, 
there is one $CP$-violating phase required to accommodate the $R_{D^{(*)}}$ anomaly, and hence
$CP$-violating phenomena are inevitably predicted.
We investigated various EDMs and found that neutron and proton EDMs are induced with  opposite signs, and  predicted magnitudes are
well 
within the reach of the sensitivities of future experiments.
It is also found that in the $U_1$ LQ scenario $B_s \to \tau^+\tau^-$ at LHCb Run\,3 will become another smoking-gun signal, while it is absent in the $R_2$ LQ scenario.

Correlations with other CPV phenomena, \eg, $\Delta A_{CP} (B\to X_s\gamma)$, 
will also be interesting and we leave them as a future work.
It is known that the remaining discrepancies in $b\to s\ell^+ \ell^-$
could  also be solved by the $U_1$ LQ 
at one-loop level \cite{Crivellin:2018yvo}.
Going beyond the leading-log approximation  is necessary for the presence of vector-like fermions, and it will also be a part of future work.

\section*{Acknowledgements}

We thank Yohei Ema, Ulrich Nierste, Shohei Okawa, and Maxim Pospelov for their valuable comments and discussions.
Furthermore, we would like to thank Hector Gisbert and Joan Ruiz Vidal for worthwhile discussions.
We also appreciate Felix Wilsch for the technical support of \texttt{HighPT}.
S.\,I. enjoys the support from the Deutsche Forschungsgemeinschaft (DFG, German Research Foundation) under grant 396021762-TRR\,257.
S.\,I. would like to appreciate the ``hot'' hospitality at Universidad de Barcelona where the last stage of this project was made. 
T.\,K. was supported by the Grant-in-Aid for Scientific Research (C) from the Ministry of Education, Culture, Sports, Science, and Technology (MEXT), Japan, No.\,21K03572.
This work is also supported by the Japan Society for the Promotion of Science (JSPS)  Core-to-Core Program, No.\,JPJSCCA20200002.

\appendix

\section{Other \texorpdfstring{\boldmath{$b\to c \tau\ov \nu$}}{b to c tau nu} observables}
\label{sec:Appendix_other_bctaunu}

\begin{figure*}[t]
\includegraphics[width=0.45\textwidth]{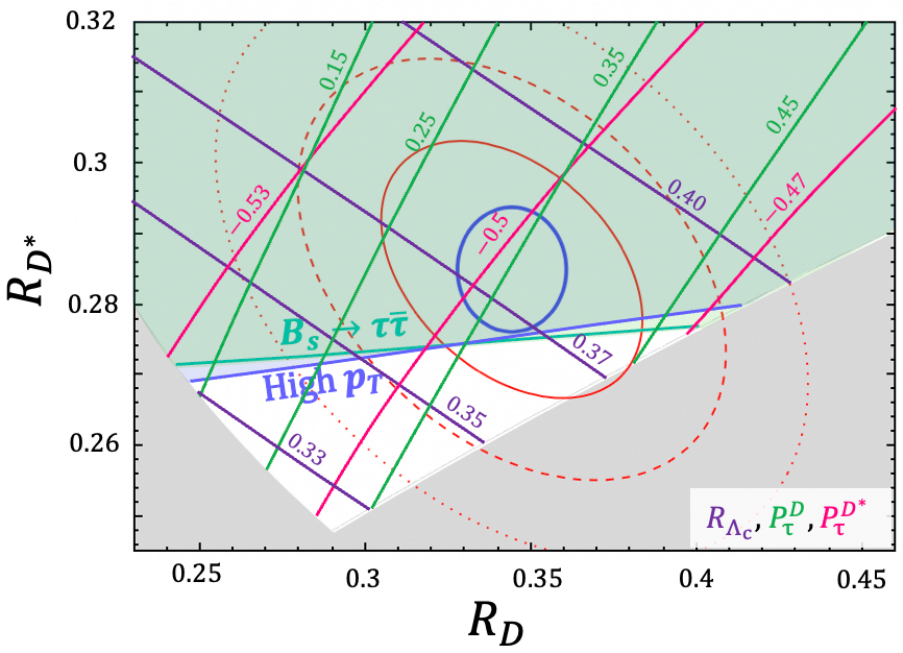}
\qquad 
\includegraphics[width=0.435\textwidth]{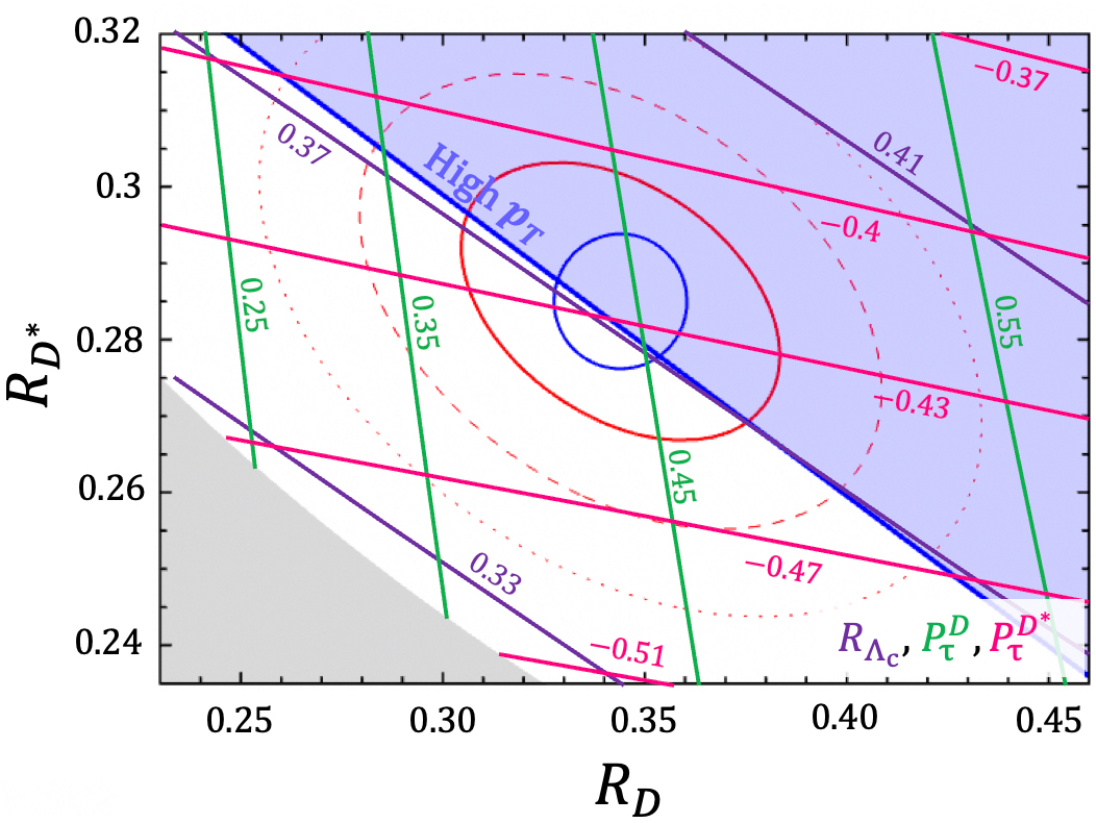}
\caption{\label{fig:polari}
Correlations with the $\tau$ polarization asymmetries, $P_\tau^D$ and $P_\tau^{D^\ast}$, and the LFU violation 
 in $\Lambda_b \to \Lambda_c l \ov\nu$, $R_{\Lambda_c}$, are shown on the plane of $R_{D}$--$R_{D^{*}}$  by the green, magenta, and purple contours, respectively, in the simplified $U_1$ (left panel) and $R_2$ LQ scenarios (right). 
}
\end{figure*}

In this Appendix, 
other related observables in $b \to c \tau \ov\nu$ are discussed in the simplified $U_1$ LQ model and the $R_2$ LQ model.
In Fig.~\ref{fig:polari}, 
the $\tau$ polarization asymmetries
in $\Bb \to D^{(\ast)}\tau \overline{\nu}$, $P_\tau^D$ and $P_\tau^{D^\ast}$  \cite{Tanaka:2012nw,Asadi:2018sym}, and the LFU violation 
 in $\Lambda_b \to \Lambda_c l \ov\nu$, $R_{\Lambda_c} \equiv \mathcal{B}(\Lambda_b\rightarrow \Lambda_c \tau \ov\nu_\tau)/ \mathcal{B}(\Lambda_b\rightarrow \Lambda_c \ell\ov\nu_\ell)$, are shown  by the green, magenta, and purple contours, respectively, for the simplified  $U_1$ and $R_2$ LQ models.

 In the $U_1$ LQ scenario, it is found that $P_\tau^{D^\ast}$ cannot deviate from the SM prediction $P^{D^\ast}_{\tau, \text{SM}}\simeq -0.50$, while $P_\tau^{D}$ can deviate from $P^{D}_{\tau, \text{SM}}\simeq 0.33$ which will be probed by the Belle~II experiment with good accuracy \cite{Alonso:2017ktd}.
On the other hand, a large value of $R_{\Lambda_c}$ is expected compared to the SM prediction, $R_{\Lambda_c}^{\rm SM}\simeq 0.32$ \cite{Bernlochner:2018kxh}. 
This behavior is consistent with a sum rule prediction \cite{Blanke:2018yud,Blanke:2019qrx,Fedele:2022iib}, and it should also be a smoking-gun signal in the LHCb experiment \cite{LHCb:2022piu}.
Note that the $D^*$ longitudinal polarization ratio
in $\Bb \to D^{\ast}\tau \overline{\nu}$, $F_L^{D^*}$ \cite{Tanaka:2012nw,Belle:2019ewo}, is also predicted. It is, however, found that the $U_1$ LQ effect is tiny, 
$\Delta F_L^{D^*}=F_L^{D^*} - F_{L, {\rm SM}}^{D^*}= \pm 0.01$ \cite{Iguro:2018vqb,Fuentes-Martin:2019mun},
and it is smaller than the Belle II sensitivity \cite{Belle-II:2018jsg}.

In the $R_2$ LQ scenario, 
both $P_\tau^D$ and $P^{D^\ast}_{\tau}$ are expected to deviate from the SM predictions \cite{Iguro:2024hyk}.
Also, the  large value of $R_{\Lambda_c}$ is expected in accordance with the sum rule.
On the other hand, it is expected that $F_L^{D^*}$  cannot deviate: 
$\Delta F_L^{D^*}\simeq - 0.01$ \cite{Iguro:2018vqb}.

\bibliographystyle{utphys28mod}
\bibliography{ref}

\end{document}